# Gamma-ray and high-energy X-ray detection with large area scintillating crystals: a hands-on review

**Maurizio Bonesini** [1,*]

[1]*Sezione INFN and Dipartimento di Fisica G. Occhialini*
*Universitá Milano Bicocca, Piazza Scienza 3, 20123, Milano, Italy*

Correspondence*:
M. Bonesini
maurizio.bonesini@mib.infn.it

## ABSTRACT

Detection of photons with scintillating inorganic crystals in the high-energy range ($>$0.1 MeV) will be discussed, making a comparison with other available methods. Energy resolutions up to 2 % at 662 keV and fast decay time of the order of 20 ns are within reach, with the introduction of Ce-doped crystals at the place of alkali-halides ones. Development are under way for the production of non-hygroscopic scintillating crystals, such as PrLuAg and Ce: GAAG. At the end of this review, examples of experimenta devices based on scintillating inorganic crystals will be discussed. Practical hands-on experience is emphasized at the expense of a more comprehensive description of all available and possible options. Detectors' construction details and consequences of the different choices will be discussed. The emphasis will be put on the LaBr$_3$: Ce based detectors that are the present "golden standard" in gamma ray spectroscopy. The focus of this review will be on photon detection in the high-energy region: mainly 0.1 - 2 MeV, including both gamma rays and high-energy X-rays, even if many considerations may be applied to the detection of low-energy X-rays.

## 1 INTRODUCTION

Photons in the high-energy region ($>$0.1 MeV) include both gamma rays from nuclear decays and high-energy X-rays, such as the ones from muonic atomic decays. In the following, we will not care about their origin: nuclear or atomic and will use the generic term gamma rays to cover both types of radiation.

X-rays were discovered about 130 years ago by W.C. Roentgen, by observing a glow on a phosphor screen (Roentgen, 1896). As direct registration of X-rays on a photographic plate is quite inefficient, a search for materials to convert X-rays to visible/ultraviolet (UV) light, to be detected later by a photographic plate, started immediately. Powders such as CWO$_4$ (Edison, 1896) or Zns-based ones were then introduced. A similar indirect detection methos is used with inorganic scintillating crystals. Direct detection of X-rays may be done instead with semiconductor detectors. Here the incoming radiation is directly converted into the output signal, without an intermediate step.

Gamma-ray detectors in common use may be divided into three main categories:

- Gas-filled detectors
- Semiconductor crystal detectors
- Inorganic scintillating crystal detectors

The choice depends on the energy range of interest, the needed energy resolution and the required detection efficiency. In addition, other requirements such as count rate performances, signal pulse shape and cost may be of relevance.

In the high-energy region (>0.1 MeV) for photon detection, scintillating inorganic crystals are the most common choice for large area detectors and they will be fully reviewed in the following.

## 2 DETECTORS BASED ON SCINTILLATING INORGANIC CRYSTALS

Common materials include Sodium Iodide (NaI(Tl)), Bismuth Germanate (BGO), Lutetium-Yttrium oxy orthosilicate (LYSO), Lanthanum Bromide (LaBr$_3$:Ce) and many others. Their great importance is associated with their good energy and time resolution, their high counting rate capability (up to $10^7$ counts/s), their high detection efficiency and finally their variety in size and making.

The main characteristics required for a scintillating crystal are:

- high material density: in the range of 3-7 g/cm$^3$, with high atomic number of the major constituent, to allow high detection efficiency for Gamma-rays (high "stopping power")
- small decay time of the crystal's fluorescent component providing fast signals, thus allowing high counting rates
- high light yield improving the photon statistics and thus the energy resolution
- small response non-linearity, giving a small degradation in energy
- chemical stability and radiation hardness
- matching of the crystal peak emission wavelength with the photodetector peak response

The first scintillating crystals used for gamma ray detection were Alkali Halide ones (such as NaI, NaCl and NaBr) with Thallium as activator (Pohl, 1938). If the Thallium concentration is small (up to 0.1 molar percent) luminescence is in the near UV region, otherwise if it is larger (from 0.1 to 5 molar percent) the emission extends to the visible region. A crystal, as NaI(Tl), is a non-conducting crystal. This means there is a large energy gap between its filled valence band and its empty conduction band. Energetic electrons generated by a gamma interaction with the NaI material will lose their kinetic energy by producing electron-hole pairs. The recombination of these pairs may result in light emission through radiative transitions or energy release as lattice vibrations. The inclusion of Thallium at a $10^{-3}$ molar fraction significantly enhances the crystal's light emission, acting as an activator. The radiative emission of Thallium follows an exponential decay law with a large decay constant, meaning that luminescence occurs within a few microseconds. This enables the distinction of different scintillation events in time, which is crucial for differentiating the timing of gamma photon energy depositions.

In a simple phosphor the number $N$ of visible/UV photons produced in the scintillation conversion by an incoming gamma ray of energy E, may be expressed by:

$$N = \frac{E}{\beta E_{gap}} \times QE_T \times QE_L$$

where $E_{gap}$ is the energy of the forbidden gap, $QE_T(QE_L)$ are the quantum efficiencies of the transport(luminescence) stages in visible/UV light production, $\beta$ a phenomenological parameter in the range 2-3. The relative conversion efficiency is thus:

$$\eta = \frac{E_{vis} \times N}{E}$$

with $E_{vis}$ energy of the generated visible/UV photons. For the best available material $\eta$ reaches a value of $\sim 0.2$. For a scintillator as an inorganic crystal, one has to take into account also the collection time of photons after the gamma ray absorption, giving at the end a lower value for the efficiency $\eta$.

After some preliminary studies on phosphors as Naphtalene (Kallman, 1947), Anthracene (Bell, 1948) and Calcium Tungstate (Moon, 1948), the use of NaI(Tl) crystals for gamma detection was introduced in seminal papers by R. Hostader (Hofstader, 1948), (Hofstader, 1949) in the late 40's. They were the crystals of election for many years having a good photon yield and a reasonable energy resolution even if with a long signal decay time ($\sim 250$ ns). The additional problem of being hygroscopic was handled by a proper Aluminum housing.

NaI(Tl) typically converts around 11 % of the incident gamma energy into photons, with an average energy of 3.0 eV per photon. For a 1 MeV photon, approximately $3.8 \times 10^4$ photons are produced on average. The statistical fluctuations in the number of photons generated by each gamma contribute to the observed width (energy resolution) of the observed photopeaks.

## 2.1 Available crystals

A selection of scintillating crystals in current use for gamma ray detection is shown in Table 1, with their main properties.

Crystals are made from compounds, with a melting point in the typical range 700 -2000 $°C$. They can be grown using melt-based methods such as the ones from Bridgman or Czochralski, These methods are suitable for growing large volume crystals (Brice, 1986).

The first relevant distinction is between hygroscopic crystals, where an encapsulation is needed, and not-hygroscopic ones. While PrLuAG crystals (Drozdowski et al., 2008) and Ce: GAAG crystals (Yeom et al., 2013) crystals are not-hygroscopic and thus do not need encapsulation, the more conventional LaBr$_3$: Ce (van Loef et al., 2001), CeBr$_3$ (Quarati et al., 2012) and NaI(Tl) scintillating crystals are hygroscopic. Their main properties are also shown in Table 1, which is a compilation from published data (Workman et al., 2022) and producers' datasheets. Reported crystal's thickness ($\Delta z$ in cm) for 88 % attenuation at lower energy is computed from mass attenuation coefficients, as reported in (Hubbell and Seltzer, 1996) and confirmed by Monte Carlo calculations with the MNCP code (Carter et al., 1975)

NaI(Tl) crystals have been recently superseded by more performant Ce-doped crystals such as LaBr$_3$:Ce, Lanthanum BromoChloride (LBC) and others, especially concerning the signal timing. At photon energies of 662 (122) keV LBC has a full width at half maximum (FWHM) energy resolution of $\sim$ 3 % ($\sim$ 6.4 %) to be compared with 7 % (9 %) obtained with a NaI(Tl) crystal. The "golden standard" LaBr$_3$: Ce has instead FWHM energy resolution of $\sim$ 2.9 % ($\sim$ 6.6 %) at 662 (122) keV.

LaBr$_3$:Ce is the crystal of choice for X-ray spectroscopy due to its high light output ($\sim 60000$ $\gamma$/MeV), fast decay time ($\tau \sim 30$ ns) and small non-linearity (less than 5 % ). This non-linearity compares well with what has been measured for Na(Tl) ( 20 % ). Its scintillation properties are connected with the used Cerium concentration, as shown in reference (Shah et al., 2002). Going from 0.5 % concentration to 5 %, the light output decreases by $\sim$ 10 %, while the decay time goes from 26 to 15 ns.

**Table 1.** Main characteristics of the crystals commonly used for X-ray detection. Typical energy resolutions (FWHM in %), from published data, are measured with a PMT readout.

| Scintillators | PrLuAG | Ce:GAGG | $LaBr_3:Ce$ | $CeBr_3$ | LBC | LYSO | Csi(Tl) | NaI(Tl) | BGO |
|---|---|---|---|---|---|---|---|---|---|
| Density ($g/cm^3$) | 6.73 | 6.63 | 5.08 | 5.18 | 4.90 | 7.20 | 4.51 | 3.67 | 7.13 |
| Light yield ($\gamma$/MeV) | 22,000 | 57,000 | 75,000 | 47000 | 57000 | 30000 | 54000 | 38000 | 10000 |
| Decay time (ns) | 20 | 88 (91 %) 258 (9 %) | 30 | 25 | 35 | 40 | 900 | 250 | 300 |
| Peak emission (nm) | 310 | 520 | 375 | 370 | 380 | 420 | 550 | 415 | 480 |
| Energy res (%) @ 662 keV | 4.3 | 5.3 | 2.9 | 4.0 | 3.0 | 8.4 | 6.5 | 7.0 | 10.0 |
| Energy res (%) @ 120 keV | - | - | 6.6 | 10.0 | 6.4 | - | 12.0 | 9.0 | 14.4 |
| Hygroscopic | no | no | yes | yes | yes | no | yes (slightly) | yes | no |
| Melting point ($^\circ C$) | 2043 | 1850 | 1116 | 722 | - | 2047 | 621 | 651 | 1044 |
| $\Delta z$ (cm): 88 % att. @ 100 keV | 0.12 | 0.18 | 0.33 | 0.31 | - | 0.12 | 0.23 | 0.35 | 0.08 |
| $\Delta z$ (cm): 88 % att. @ 200 keV | 0.65 | 0.91 | 1.54 | 1.48 | - | 0.67 | 1.25 | 1.76 | 0.43 |

All inorganic crystals based on Lanthanum or Lutetium suffer from an intrinsic activity, due either to the presence of the $^{176}$Lu isotope in naturally occurring lutetium [1] or $^{138}$La which emits conversion electrons and $\beta$ particles with energy up to 1.7 MeV (Bonesini et al., 2016), (Iyudin et al., 2009). While the intrinsic activity of PrLuAg crystals is not negligible ($\sim$ 36 Bq/g), the intrinsic activity of Ce: GAAG crystals is minimal ($\leq 1.5 \times 10^{-3}$ Bq/g). The activity of $LaBr_3$:Ce ($\sim$ 0.2 Bq/g) is half way between the two.

$CeBr_3$ scintillating crystals (Quarati et al., 2007), (Fraile, 2013), (Ackerman, 2015) offer an alternative to NaI(Tl) for high resolution gamma ray spectroscopy. With FWHM energy resolution similar to the one of $LaBr_3$:Ce, they do not suffer from the $^{138}$La background typical for La-halide crystals, such as $LaBr_3$: Ce and LBC. However, they have a small intrinsic background due to $^{227}$Ac, giving a number of peaks between 1500 and 2200 keV.

The new Lanthanum BromoChloride (LBC) crystals have similar properties to $LaBr_3$: Ce ones, but are mechanically stronger. With similar resolution at the $^{137}$Cs peak, LBC suffers from the same $^{138}$La problems as LaBr3:Ce.

For a more complete review of the available scintillating inorganic crystals, the reader may use references (Nikl, 2006) and (Yanagida, 2018).

Photon detectors in the high energy range ($\geq$ 0.1 MeV) have applications in many fields. Examples are TOF PET imaging (Moses and Shah, 2005), fundamental physics such as the FAMU measurement at Riken-RAL of the Zemach proton radius (Pizzolotto et al., 2020), gamma-ray astronomy (Gostojic et al., 2016) and homeland security (Zentai, 2008).

---

[1] $\tau_{1/2} = 3.78 \cdot 10^{10}$ years, 2.59 % abundance

## 2.2 Assembly of crystal based detectors

Crystals are made commonly in cylindrical shape or in cubic shape. Usually, only one surface of the scintillator is designed to provide the light output, while the others are coated with diffusive or reflecting materials [2]. This introduces the problem of matching the terminal face or the optical window of the crystal with the surface of the readout device. If SiPM arrays are used, nearly circular sizes may be obtained only by a custom mounting, as done for example in the FATIMA experiment (Pascu et al., 2024), using $3 \times 3$ and $6 \times 6$ mm$^2$ SiPM. In the standard mounting, either cubic crystals are used or some active area of the SiPM array does not see the crystal end-face. This implies a moderate increase in the dark noise. The optical coupling between crystal and the photodetector face is usually obtained by optical glue or Silicon optical grease. More details are reported in the following. Detectors based on crystals may be mounted in different configurations, according to experimental requirements, even if standard cylindrical or parallelepiped form mounting are available from producers such as Berkeley Nucleonics, Bicron, CAEN and Nuclear Instruments. An example of a custom detector's mounting from the FAMU experiment at RAL is shown in the left and middle panels of Figure 1. The crystal holder was made in ABS with a 3D printer.

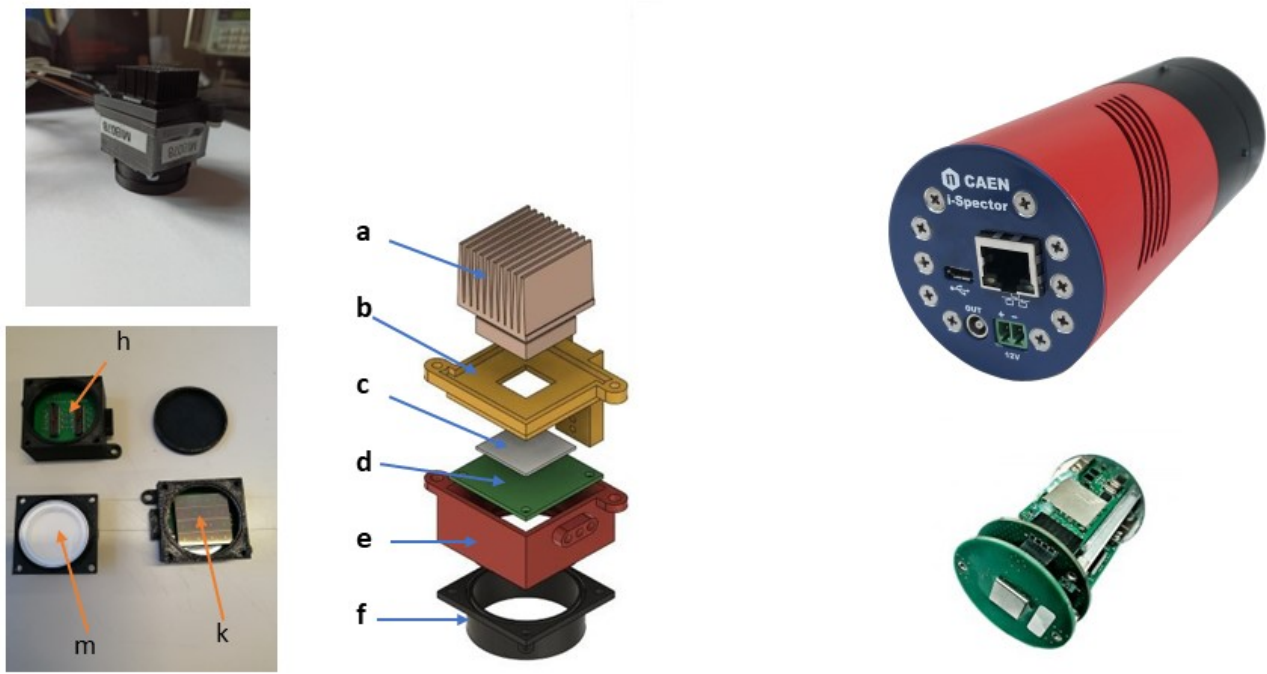

**Figure 1.** **Left panel**. Top-left: image of a complete 1" X-ray detector for the FAMU experiment at RAL. Bottom-left: images of some details of the crystal holder: h) with the Printed Circuit Board (PCB) inside, k) with mounted SiPM array, m) with crystal inside. Right : exploded view of a 1" detector. From top to bottom: a) heat dissipator, b) detector base; c) gap filler, d) PCB , e) PCB holder, f) crystal holder. **Right panel**. Top-right: image of the i-Spector detector. Bottom-right: exploded inside view of one i-Spector detector. Crystals of different types and sizes may be provided (courtesy of CAEN srl).

An innovative off-the-shelf solution is instead the i-Spector detector from CAEN, shown in the right panel of figure 1. It is a fully integrated tube designed to replace existing systems based on PMTs. It includes a SiPM array, an amplifier stage, an integrated power supply for biasing of SiPM and temperature drift correction of SiPM gain.

---

[2] For crystals emitting around 310 nm, as PrLuAg, it is difficult to find a proper optical diffuser. The choice in reference (Bonesini et al., 2016) was the Avian-B optical coating, based on BaSO$_4$. A reflectance $\geq 97\%(\geq 92\%)$ is quoted for it in the range 350-850 nm (250-1300 nm)

### 2.2.1 Optical coupling between scintillating crystals and photodetectors

When the bottom face of a crystal has a similar or equal area to the photodetector's one a simple coupling, based on optical glue or optical cement, may be used [3]. If instead, there is a large mismatch in the two areas a light guide is utilised. Such light guides are made of optical quality plexiglass, lucite or perspex and work on the principle of internal reflection. Light is "guided" from one end to the other via internal reflections between the external walls of the light guide. For this scope the external walls are polished. As the given flux of light at the input can never be concentrated into a smaller cross-sectional area at the output (Garwin, 1970), other methods, such as Winston cones (Winston, 1970) have to be used, to maximize the collection of incoming rays. A Winston cone is an off-axis parabola of revolution designed to maximize collection of incoming rays within some field of view. Winston cones are nonimaging light concentrators intended to funnel all wavelengths passing through the entrance aperture out through the exit aperture. They maximize the collection of incoming rays by allowing off-axis rays to make multiple bounces before passing out the exit aperture.

## 2.3 Readout techniques

The output signal from scintillating inorganic crystals may be read by different photodetectors such as photomultipliers (PMTs), Silicon Photomultipliers (SiPMs), Silicon Avalanche Photodiodes (Si-APD), Silicon Drift Detectors (SDDs). The advantages and disadvantages of the different solutions will be discussed in the following. Their main characteristics are resumed in Table 2, where Q.E. is the peak Quantum Efficiency and $V_{op}$ the operating voltage.

**Table 2.** Main characteristics of used photodetectors for crystal readout. Typical values are reported.

|  | PMTs | SiPMs | SiPM Arrays | Si-APD | SDDs |
|---|---|---|---|---|---|
| $V_{op}$ | 1000-2000 V | 25-70 V | 25-70 V | 150-400 V | 100 V |
| **B** sensitive | yes | no | no | no | no |
| signal sign | -ve | +ve | +ve | +ve | +ve |
| gain (typical) | $10^6$ | $10^6$ | $10^6$ | 50-100 | 1 |
| Q.E. (peak) | 25-40 % | 30-50 % | 30-50 % | 67 % | 70-85 % |
| max size | up to 3" round up to 1" square | up to 6x6 mm$^2$ | up to 1" square | up to $10 \times 10$ mm$^2$ | up to $8 \times 8$ mm$^2$ |

### 2.3.1 PMT's based readout

A PMT has been the conventional choice for many years for crystal readout. The main characteristics of some PMTs from Hamamatsu Photonics or Photonis, currently in use, are shown in table 3. When a good energy resolution is needed, PMTs with higher photocathode quantum efficiency (QE) have to be used. Hamamatsu has recently introduced an Ultra BiAlkali (UBA) photocathode (QE $\simeq 42\%$ at 380 nm) replacing the old BiAlkali (BA) one (QE $\simeq 22\%$ at 380 nm). The use of high photon yield crystals coupled to high-efficiency photocathodes may produce very high peak currents with dynodes current saturation, producing non linear effects. To deal with these effects, one may reduce the number of dynodes or use a tapered voltage divider. In this last design either the voltage gradient is enhanced in the first and/or last few stages. The output linearity is thus improved. Another way to increase the linearity is to use transistors or Zener diodes instead of resistors in the last few stages of the divider. This active divider ensures a good linearity up to an output current of $\sim 60-70$ % of the voltage divider current, as explained in (Hamamatsu, 2017). At low detector rates ($\leq 10$ kHz) no difference is seen between resistor-type and active-type voltage divider, as shown in reference (Gandolfo et al., 2023).

---

[3] Bicron BC600 (BC630) is a typical optical cement (grease) used in this case. BC630 has a refractive index of 1.465

**Table 3.** Main characteristics of some PMTs from Hamamatsu or Photonis in common use for crystal readout.

|              | PMT type | no of dynodes | Q.E. (%) | photocathode type | size | typ gain (x $10^6$) |
|---|---|---|---|---|---|---|
| R6231-01     | standard | 8  | 30 | BA  | 51 mm round       | 0.27 |
| R7600-200    | standard | 10 | 44 | UBA | 18 x 18 mm$^2$    | 1.3  |
| R7600        | standard | 10 | 22 | BA  | 18 x 18 mm$^2$    | 2.0  |
| H8500C       | MA-PMT   | 12 | 27 | BA  | 49 x 49 mm$^2$    | 2    |
| R9420        | standard | 8  | 27 | BA  | 38 mm round       | 0.50 |
| R11265U-200  | standard | 12 | 43 | UBA | 23 x 23 mm$^2$    | 0.35 |
| R6233-01     | standard | 8  | 30 | BA  | 76 mm round       | 0.27 |
| R10233-100   | standard | 8  | 35 | BA  | 76 mm round       | 0.23 |
| XP5200       | standard | 8  | 30 | BA  | 51 mm round       | 2.4  |

Examples of LaBr$_3$:Ce detectors with a PMT readout are reported in references (Pani et al., 2008), (Omer et al., 2013) (Cinti et al., 2013), (Gandolfo et al., 2023), (Baldazzi et al., 2017), (Chewpraditke and Moszynski, 2011), (Quarati et al., 2007), (Swiderski et al., 2015), (Giaz et al., 2014) and (Giaz et al., 2013). Their main characteristics are shown in Table 4.

**Table 4.** Main characteristics of some LaBr3:Ce detectors with PMT readout. R is the FWHM energy resolution.

|  | PMT type | R( % ) |
|---|---|---|
| 1/2" round, 1/2" thick (Pani et al., 2008) | R7600-200 | 3.2 @ 662 keV |
| 1.5" round, 3" thick (Omer et al., 2013) | R9420 | 2.65 @ 847 keV |
| 51 x 51 mm$^2$, 4 mm thick Cinti et al. (2013) | H8500C-100 | 4.11 @ 511 keV |
| 1.5" round, 2" thick (Gandolfo et al., 2023) | R6231 | 3.1 @ 662 keV |
| 1" round, 1" thick (Baldazzi et al., 2017) | R11265U | 3.5 @ 662 keV |
| 1/2" round, 1/2" thick (Chewpraditke and Moszynski, 2011) | XP5500B | 3.5 @ 662 keV |
| 2" round, 2" thick (Quarati et al., 2007) | R6231 | 3.0 @ 662 keV |
| 1" round, 1" thick (Swiderski et al., 2015) | XP5200 | 3.0 @ 662 keV |
| 3" round, 3" thick (Giaz et al., 2014) | R6233 H8500 | 3.2 @ 662 keV |
| 3.5" round, 8" thick (Giaz et al., 2013) | R10233-100 | 3.1 @ 662 keV |

While LaBr$_3$:Ce crystals of small size were grown starting from 2001, only in 2008 Saint Gobain Crystals was able to grow large size crystals up to 3.5" x 8", under test in reference (Giaz et al., 2013). For these large size crystals, performances are affected in addition by self absorption, possible internal non-homogeneity that may affect the light yield and the longer mean free path to the photodetector's front face. As a consequence their properties may not immediately extrapolated from the ones of smaller crystals.

### 2.3.2 SiPM based readout

Silicon photomultipliers (SiPM) are a valuable alternative to conventional photomultipliers (PMTs) for the readout of scintillation detectors [4]. As readout devices, they have high reliability, low sensitivity to external magnetic fields and can operate at voltages significantly lower than the ones used for PMTs. Using a SiPM array for the readout of a inorganic scintillating crystal it is possible to obtain energy resolutions comparable to what is obtained with PMTs. However, SIPMs have a relevant problem: in addition to an increased noise level, their gain drifts significantly as a function of temperature. This feature prevents their use in conditions with a changing temperature, as homeland security and military applications, unless an ad-hoc correction: offline or online, is implemented.

To read large area crystals instead of single SiPM (max area $6 \times 6$ mm$^2$) SiPM square arrays of typical size 1" or 1/2" are used. A larger area (up to 2") may be obtained by combining single SiPMs, as done in reference (Du et al., 2016), at the cost of increased dark noise and the need to engineer a custom PCB. The main characteristics of some available SiPM arrays are reported in Table 5. Operating voltages $V_{op}$ are set to $V_{brk} + \Delta V$, where the overvoltage $\Delta V$ is chosen according to the manufacturer's specifications. Typical overvoltages are in the range 2-5 V.

**Table 5.** Main characteristics of common SiPM 1/2 " and 1" arrays. Data are from producers' datasheets

|  | size (inches) | cell dim (mm$^2$) | $V_{op}$ (V) | $\Delta V_{bd}/T$ mV/C | $\lambda_{peak}$ (nm) | PDE max (%) | spectral range (nm) |
|---|---|---|---|---|---|---|---|
| Hamamatsu S14161-6050-AS | 1 | 6x6 | 41.1 | 34 | 450 | 50 | 270-900 |
| SENSL Array-J-60035-4P | 1/2 | 6x6 | 29 | 21.5 | 420 | 50 | 200-900 |
| Advansid NUV3S-4x4-TD | 1/2 | 3x3 | 29.5 | 26 | 420 | 43 | 350-900 |
| Hamamatsu S14161-3050-AS | 1/2 | 3x3 | 41.1 | 34 | 450 | 50 | 270-900 |
| Hamamatsu S13161-3050-AS | 1/2 | 3x3 | 53.8 | 60 | 450 | 35 | 320-900 |
| SENSL SB-4-3035-CER | 1/2 | 3x3 | 26 | 21.5 | 420 | 30 | 300-800 |
| Advansid NUV3S-4x4TD | 1/2 | 3x3 | 29 | 26 | 420 | 43 | 350-900 |
| Broadcom AFBR-S4N | 1/2 | 3x3 | 40 | 14.6 | 420 | 63 | 250-900 |

The Photon Detection Efficiency (PDE) of SiPM arrays for the light of different wavelength depends also on the type of window used: epoxy or Silicone. A more fragile silicone window is used to increase the response around 380 nm (near UV) to match the light emission of LaBr$_3$:Ce or PrLuAg crystals. For other types of crystals epoxy type windows, with better mechanical characteristics, are to be preferred.

In the case when the signals from the SiPM cells in the readout of a SiPM array are summed up, the different cells may be powered using different schemes ("ganging"). The type of "ganging" used has a relevant influence on the shaping time of the signal (especially the fall time).

---

[4] A SiPM is a set of miniature avalanche photodiodes operating in Geiger mode, connected in parallel. Their outputs are connected to one common output. For a full discussion see references (Buzhan et al., 2003) and (van Dam et al., 2010).

A simple processing scheme, based on a Flash Analog to Digital Converter (FADC), digitizes the input signal, producing both charge, amplitude and timing informations. This scheme compares well with the one used in a standard spectroscopic chain where the input signal is shaped by a spectroscopic amplifier [5] going to a Multi-Channel Analyzer (MCA). In the following, results with different ganging schemes will be shown, going from standard parallel ganging to hybrid ganging and finally to a custom 4-1 scheme developed by Nuclear Instruments.

Examples of LaBr$_3$:Ce detectors with a SiPM Array readout are reported in references (Vita et al., 2022), (Poleshchuck et al., 2021), (Bonesini et al., 2023b), (Bonesini et al., 2023a), (Cozzi et al., 2017) and (He et al., 2023). Their main characteristics are shown in table 6.

**Table 6.** Main characteristics of some LaBr$_3$: Ce detectors with SiPM array readout. R is the FWHM energy resolution.

|  | SiPM type | R (%) @ 662 keV |
|---|---|---|
| 3" round, 3' thick (Vita et al., 2022) | FBK NUV-HD | 2.6 % |
| 1.5" cubic (Poleshchuck et al., 2021) | SENSL ArrayJ-60035 | 2.94 % |
| 1/2" cubic, 1/2" thick (Bonesini et al., 2023b) | S13161-3050 | 3.27 % |
| 1" round, 1" thick (Bonesini et al., 2023a) | S14161-6050 | 3.01 % |
| 2" round, 2" thick (Cozzi et al., 2017) | FBK NUV-HD | 3.2 % |
| 3" round, 15 mm thick (He et al., 2023) | SENSL 60035-TVS | 5.3 % |

### 2.3.2.1  *Correction of SiPM gain drift with temperature*

The breakdown voltage $V_{bd}$ of a SiPM varies with temperature according to the following equation:

$$V_{bd}(T) = V_{bd}(T_0) \times (1 + \beta(T - T_0))$$

where T$_0$ is the reference temperature (typically 25 °C), and $\beta$ is the temperature coefficient of the SiPM, given by $\frac{\Delta V_{bd}}{\Delta T}$. As a result, the operating voltage $V_{op} = V_{bd} + \Delta V$, where $\Delta$V is the overvoltage, must be adjusted accordingly to maintain a consistent gain, as discussed in references (Dinu et al., 2010) and (Otte et al., 2017). The response of a typical 1" detector to a $^{137}$Cs source in a Memmert IPV-30 climatic chamber, where the temperature spans the range between 20 and 30 °C, is shown in the top panel of Figure 2. Without the online temperature correction, the resolution of the 662 keV photopeak is significantly degraded. With the online temperature correction applied (further details provided below), no major degradation of the $^{137}$Cs photopeak is observed.

The SiPM gain drift may be corrected by measuring the temperature of the SiPM array and making either an online or an offline correction. In addition to custom solutions, as implemented in references (Kaplan, 2009), (Eigen, 2019), (Shim et al., 2021), (Vita et al., 2022), electronic circuits for SiPM biasing and corrections with temperature are commercially available, as is the A7585 chip from CAEN. In the FAMU experiment at RAL, a solution based on the custom assembly of these chips in an 8-channels NIM module was developed. The temperature T is measured on the back side of the SiPM arrays via Analog Devices TMP37 thermistors, to correct online the operating voltage (see references (Bonesini et al., 2016), (Bonesini et al., 2022) for more details). As shown in figure 2 the effect on the detector response (pulse height (P.H.) of the $^{137}$Cs photopeak in a.u.) between 10 °C and 30 °C is reduced from 41 % to 5 % for 1"

---

[5] The Ortec 672 NIM module is a well known example

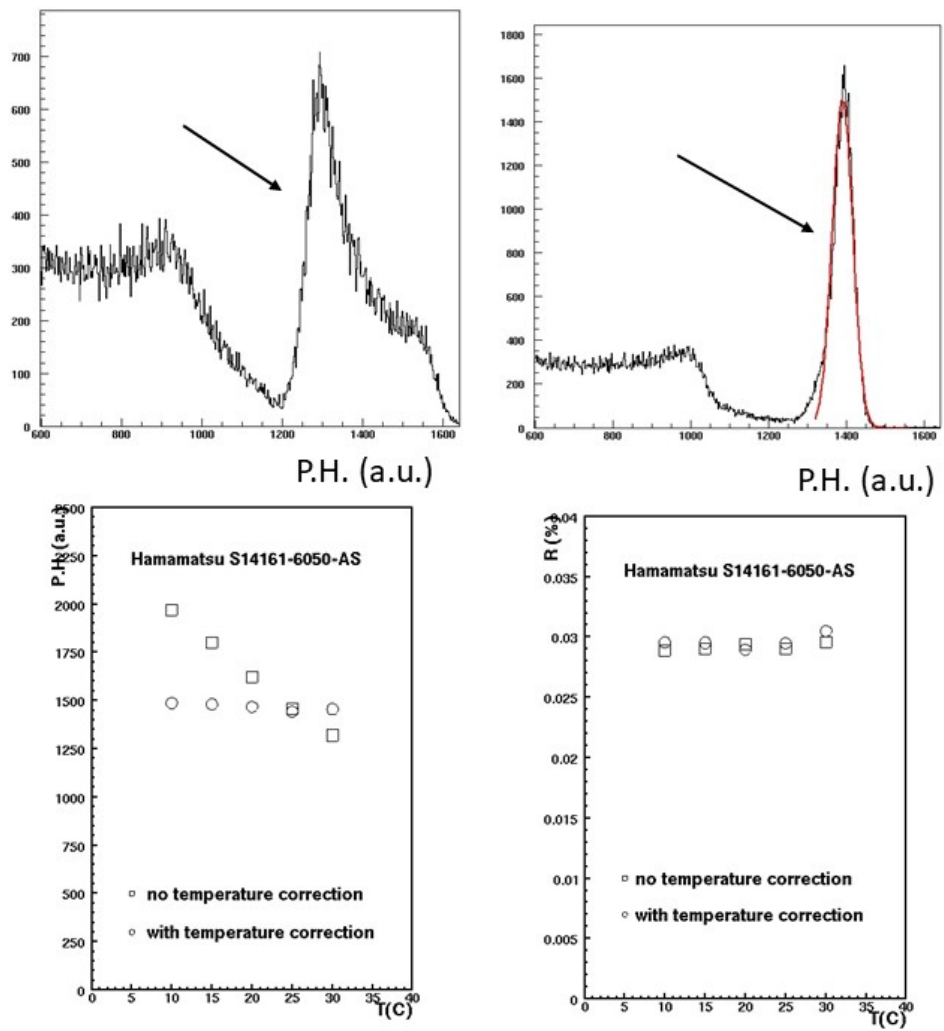

**Figure 2. Top panel**. Top-left: $^{137}$Cs spectra recorded by a LaBr$_3$: Ce 1" detector read by a Hamamatsu 14461 SiPM array during a temperature scan between 20 °C and 30 °C, inside a climatic chamber, from reference (Bonesini et al., 2022) without temperature correction. Top-right: the same with online temperature correction. Arrows point to the position of the $^{137}$Cs photopeak. **Bottom panel**. Bottom-left: dependence of the photo peak position at 662 keV for a typical 1" detector, with and without temperature correction. Bottom-right panel: dependence of the FHWM energy resolution for the same typical 1" detector with and without temperature correction.

LaBr$_3$ : Ce detectors. The custom module has an interface with the control PC based on the I2C protocol, via an FDTI USB-I2C or Arduino module.

### 2.3.2.2 SiPM ganging schemes

Single SiPMs within a SiPM array can be interconnected in various configurations depending on specific requirements, such as speed, signal-to-noise ratio (S/N), and granularity. The top panel of Figure 3 illustrates the different possible conventional configurations.

In parallel ganging, the increased capacitance results in slower rise times and longer fall times. Additionally, SiPMs with the same operating voltage $V_{op}$ must be grouped together. On the other hand, in series ganging, the charge and amplitude are reduced, leading to faster signals but requiring higher operating voltages: specifically, a factor of N more, where N is the number of individual SiPMs.

Hybrid ganging combines both series and parallel connections: the SiPMs are connected in series for the signal and in parallel for the bias, with decoupling capacitors placed between them. This configuration, originally developed for the MEG II upgrade (Ogawa, 2016), uses a common bias voltage for all the SiPMs.

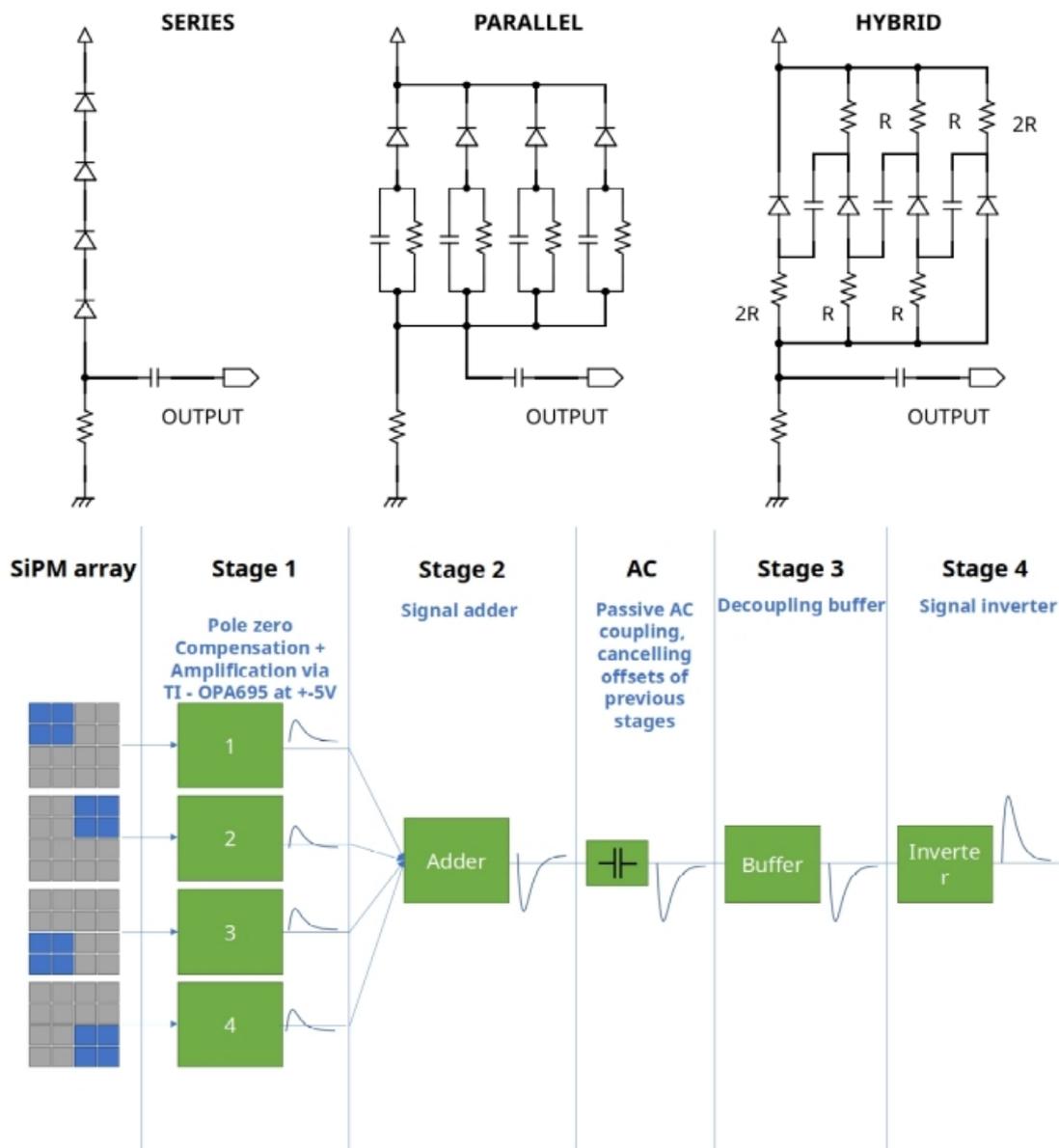

Figure 3. Top panel: layout of different ganging schemes for SiPMs: series ganging, parallel ganging, hybrid ganging, from reference (Bonesini et al., 2023b). Bottom panel: Schematic behavior of the various stages of the Nuclear Instruments 4-1 PCB circuit.

Taking into account the shape of the produced signal waveforms with the different ganging schemes, the pulse height is nearly equivalent with either a series or a hybrid ganging, while it is smaller with a parallel ganging. Time constants are bigger instead with parallel ganging and shorter with either series or hybrid ganging.

The 4-1 Nuclear Instruments circuit is based on the idea of dividing 1" square SiPM arrays into four sub-arrays, to reduce the capacitance of the single elements, dealing with smaller detectors. In a single sub-array the ganging is still parallel. As shown in the bottom panel of Figure 3 in the initial stage (stage 1) the signal from each sub-array has a pole-zero compensation stage, followed by an amplification stage via Texas Instruments OPA695 amplifiers. Signals are then added in stage 2. The following stages realize an AC coupling (to cancel offsets) and invert the output signal. For construction details, see references (Bonesini et al., 2023a) and (Bonesini et al., 2023b). Timing and energy resolution results for a typical 1" $LaBr_3$: Ce detector are shown in Table 7 for different "ganging" schemes, applied to the same 1" $LaBr_3$: Ce detector. Both with the hybrid ganging solution and the one with pole zero suppression + increased SiPM overvoltage, to compensate for signal reduction, good timing may be obtained. Unfortunately, a good FWHM energy resolution may be achieved only with the pole zero suppression + increased SiPM overvoltage($V_{over}$) solution at the expense of increased noise. For more details see reference (Bonesini et al., 2022). An optimal compromise is obtained with the 4-1 Nuclear Instruments solution, where at nominal $V_{op}$ the rise time (fall time) of the signal is reduced by a factor $\sim 2$, with respect to parallel ganging, while keeping the same good FWHM energy resolution.

Table 7. Results for a typical 1" detector, with different ganging.

|  | $V_{op}$ (V) | rise time (ns) | fall time (ns) | resolution (%) @ $^{57}$Co | resolution (%) @ $^{137}$Cs |
|---|---|---|---|---|---|
| parallel ganging | 40.82 | $68.9 \pm 7.8$ | $293.3 \pm 43.4$ | 7.78 | 2.96 |
| hybrid ganging | 41.82 | $16.1 \pm 2.4$ | $176.8 \pm 29.0$ | 9.58 | 6.08 |
| zero pole + increased $V_{over}$ | 43.02 | $58.2 \pm 15.6$ | $123.4 \pm 21.7$ | - | 2.99 |
| NI 4-1 circuit | 40.82 | $28.4 \pm 4.5$ | $140.6 \pm 21.7$ | 7.89 | 2.98 |

Studies are under way to reduce further the fall time of the 4-1 Nuclear Instruments circuit solution. An additional factor two is expected.

### 2.3.3 Alternatives readout schemes: Si-APD or Silicon Drift Detectors

Innovative readout schemes for crystals, based on Si-APD or Silicon Drift Detectors (SSD) have been recently proposed, but their use is still quite limited.

SSD were invented in 1964 by E. Gatti and P. Rehak (Gatti and Rehak, 1984). They have a lower noise and thus a better energy resolution in principle, as compared to PMTs, smaller mass and lower power consumption, thus well fitting space applications. Being sensitive also to visible light, they may be used for the readout of scintillating inorganic crystals. As an example, in reference (Gangemi, 2016) a $LaBr_3$: Ce crystal: 0.5 " round, 0.5 " thick was coupled to a SSD developed by FBK Trento for the INFN-ASI RedSoX collaboration. A FWHM energy resolution of 3.45 % was obtained at 662 keV. The result compares well with the ones obtained with a PMT readout and may improved by a more efficient coupling between the crystal and the SSD and by an electronics with smaller noise. The authors estimate that these effects may contribute a term $\sim 1.9$ % to the measured energy resolution.

The good QE ($\sim 60$ %) obtained with the newest Si-APD, Hamamatsu S8664-55, at the LaBr$_3$: Ce emission peak, has prompted the use of these devices. Unfortunately the limited size of the available Si-APD as compared to one of the used crystals has produced worse energy resolution values, due to the poor light sampling. In reference (Scafe et al., 2007) values of 23.1 % (7.3 %) at the $^{57}$Co peak and 7.3 % (3.3 %) at the $^{137}$Cs peak are reported for a 0.5" round, 0.5" thick crystal, for a Si-APD (PMT) readout.

Using smaller LaBr$_3$: Ce round crystals of 6 mm diameter and 6 mm thickness, different readout schemes were compared, avoiding the size mismatch between crystal and detector, in reference (Moszynski et al., 2008). Results show that Si-APD and PMTs are the best solutions below 100 keV, while SSDs are better at higher energies ( $\geq 300$ keV). Clearly with larger crystals (size 1" or bigger) the light sampling problem is dominant.

## 2.4 Electronic processing chains

To process the analog signal from a crystal detector, different front-end schemes may be used. In the most simple case, a spectroscopy shaping amplifier is used. The amplifier is then followed by a MCA. If the analog signal is sizeable it may be instead fed directly into a Flash to Analog Converter (FADC) channel as shown in reference (Bonesini et al., 2016). For segmented crystal detectors, many engineered processing chips are available, starting from the SPIROC application specific integrated circuit (ASIC) developed by the OMEGA group in 2007 (Bouchet et al., 2007). With a large dynamic range and variable gain adjustment, it digitized the input information via a 12-bit Wilkinson ADC with a conversion time of $80 \mu$s. More modern readout chips for SiPM or PMTs are commercially available from Weeroc and are described in reference (Ahmad et al., 2021).

## 3 PERFORMANCES OF SCINTILATING CRYSTAL-BASED DETECTORS

Performances of a crystal based detector involve FWHM energy resolution, linearity of the response vs impinging X-ray energies and signal timing properties as main properties. Detection efficiency, mainly connected with crystal thickness and density, has also to be considered. These properties depend on the crystal type and also on the chosen readout scheme. The energy resolution may be written as:

$$(\Delta E/E)^2 = \delta_{scint}^2 + \delta_{tr}^2 + \delta_{stat}^2 + \delta_{noise}^2$$

where $\delta_{scint}$ is the intrinsic crystal resolution, $\delta_{tr}$ is the transfer component, $\delta_{stat}$ is the statistical contribution of the readout device and $\delta_{noise}$ the dark noise contribution connected to the detector's current and the noise of the electronics. This last term is negligible with a PMT readout. The statistical contribution is given by:

$$\delta_{stat} = 2.355 \times \frac{1}{\sqrt{N_{pe}}} \times F$$

where $N_{pe}$ is the number of photoelectrons and F is the excess noise factor for SiPM or APD's or a term expressed by $\sqrt{1+\epsilon}$ for PMTs, where $\epsilon$ is the variance of the electron multiplication gain in the device (Moszynski et al., 2002) [6]. The transfer component $\delta_{tr}$ is given instead by the variance associated with coupling between crystal and photocathode. The intrinsic resolution $\delta_{scint}$ depends mainly on the non-linearity of the scintillator response (Dorenbos et al., 1995), (Moszynski et al., 2002). Other effects, such as the scintillator inhomogeneity or non-uniformity of the reflecting cover of the crystal, may also contribute. The number of photoelectrons ($N_{pe}$) is proportional to the PDE, which may be expressed for a SiPM as:

$$PDE = QE \times FF \times THR$$

where QE is the Quantum Efficiency of the photocathode, FF is the filling-factor giving the ratio of the photodetector's active area to the total area and THR the probability of electrons and holes to start the Geiger breakdown. The threshold THR depends on the applied voltage.

---

[6] for good PMTs used in gamma spectroscopy $\epsilon$ is around 0.1

Detectors' linearity and FWHM energy resolution may be studied in laboratory with calibrated radioactive sources, as $^{137}$Cs, or with X-ray machines. The linearity and the energy resolution for three common inorganic scintillating crystal detectors with a SiPM array readout are shown in figure 4. The resolution of LBC is slightly better at low energies as respect to the one of LaBr$_3$: Ce. The solid line is a fit to the data, intended to guide the eye for the LaBr$_3$: Ce crystal. The FWHM decreases linearly as a function of $\frac{1}{\sqrt{E}}$.

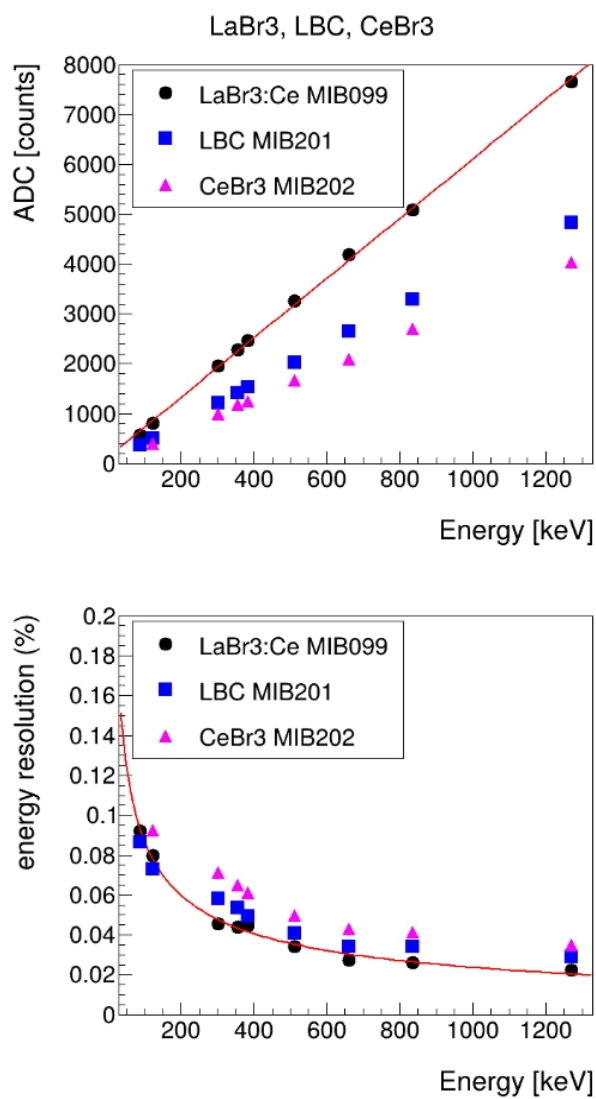

**Figure 4.** Top panel: linearity for typical 1" LaBr3:Ce, CeBr3 and LBC crystals read by an Hamamtsu S14161-6050-AS SiPM array. Bottom panel: FWHM energy resolution vs energy for the same detectors. Energy in keV refers to the incoming photon. Data were obtained in laboratory with a readout based on a CAEN V1730 fast FADC. Statistical errors are not reported, being smaller than the symbols' size.

The obtained energy resolution sometimes reduces the effectiveness of the application. This is due also to the limited light collection efficiency. As an example, in a BGO crystal, $\sim 1$ % of the emitted photons are absorbed over a 1" path and $\sim 5$ % are absorbed in the bounces between the reflective sides. The mismatch in refractive index between crystal (2.15), optical glass window (1.48) and silicon grease (1.4) produces an additional factor. At the end, only a 30-40 % light collection efficiency may be expected. To improve the energy resolution exotic proposals, such as to include the scintillating crystal inside the PMT vacuum housing and deposit the cathode directly on the scintillator surface (Chen and Belbot, 2005) were studied. In this way an increase in resolution for a BGO detector from 10 % to 6 % at 662 keV may be expected. Similar ideas were also proposed in reference (Grimm et al., 2003).

A relevant issue for crystal based detectors is their timing properties. They have a relevant impact on Positron Electron Tomography (PET) where adding time-of-flight (TOF) information enhance image to noise properties (Kuhn et al., 2006) or in experiments, as FAMU, where a fast signal fall time may enhance the signal (prompt) background (delayed) X-rays separation (Bonesini et al., 2023a). In addition a high-rate capability is a must in other experiments as NUMEN (Cappuzzello et al., 2023). The light emission from an inorganic crystal normally follows an exponential decay law:

$$I(t) \sim e^{-t/\tau}$$

with $\tau$ decay time ($\sim 30$ ns for LaBr$_3$: Ce). This simple exponential decay may be complicated by a persistence ("afterglow") corresponding to a non-exponential component, on a few ms timescale after the primary excitation has stopped (Nikl et al., 1996).

Timing issues are not a problem with a PMT based readout, where a fast PMT adds a little to the scintillator decay time. In reference (D.R.Schaart et al., 2010) a $3 \times 3 \times 5$ mm$^3$ LaBr$_3$: Ce(5 %) is read by a SiPM obtaining a 10-90 % risetime of $\sim 9$ ns and a $\sim 120$ ns falltime, to be compared to what obtained in reference (Kuhn et al., 2006) with a $4 \times 4 \times 30$ mm$^3$ LaBr$_3$:Ce(5 %) read by a Hamamatsu R4998 PMT where a 3 ns risetime is obtained. The difference may be reduced by using SiPM with a bigger fill factor and increasing the doping of the crystal with Ce: up to 30 %. The FWHM coincidence resolving time (CRT), using two LaBr$_3$:Ce detectors and one interposed Na$^{22}$ source is $101 \pm 2$ ps, corresponding to a position resolution of $\sim 15$ mm. These results may be compared with what is obtained with other small crystals coupled to SiPM. With two $3 \times 3 \times 10$ mm$^3$ LYSO: Ce crystals a CRT of 268 ps was obtained in reference (Burr and Wang, 2007) and of 240 ps in reference (Kim et al., 2009). Compared to LSO: Ce and similar materials LaBr$_3$: Ce crystals have the problem of a lower stopping power, thus requiring thicker detectors to have the same detection efficiency.

The use of larger crystals pose more severe problems due to the longer time-walk due to the increased variation of the photon path length with the interaction point (Moses, 2007). Without corrections an increase of rise time up to $\sim 29$ ns for a 4x4 SiPM 1" array is observed, as compared to $14 \pm 1$ ns for a PMT readout (Pizzolotto et al., 2020). This issue was not considered a problem in the FAMU experiment, as the relevant point was to have a signal fall time below 600 ns to be in a condition to distinguish the X-ray signal from the background.

## 3.1 Comparison with other detector types

The main alternatives to scintillating inorganic crystals are semiconductor detectors. They are based on crystal materials with a few eV band gap. Their operation is based on the direct collection of the charge carriers produced in the intrinsic region of the detector by photon interaction, applying a suitable bias voltage. Their advantage is due to a much better energy resolution. The better intrinsic resolution is due to their small Fano factor and the much smaller ionization energy required: a factor ten smaller than the one of scintillator detectors. As an example, at 1173 keV the energy FWHM resolution is $\sim 75$ keV for a NaI(Tl) scintillator and $\sim 2.35$ keV for a High Purity Germanium (HPGe) detector.

The drawback of the choice of a HPGe, as detector, is the need of cooling the Germanium crystal to reduce the intrinsic noise. At 20 $°C$ a 1 cm$^3$ sample of Germanium generates $2.5 \times 10^{13}$ electron-holes pairs from thermal energy, to be compared with a signal of $3 \times 10^5$ electron-holes pairs from a 1 MeV photon's total absorption. Thus at room temperature the signal to noise (S/N) ratio would be quite low. Cooling at cryogenic temperatures is thus mandatory to reduce the thermal noise, that manifests itself as a reverse leakage current.

While semiconductor detectors, such as HPGe, have superior energy resolution, scintillating crystal-based detectors have better timing performances. Figure 5 from reference (Bonesini et al., 2019) compares typical muonic X-ray spectra taken in the FAMU experiment at RAL, with different types of detectors. The better energy resolution of the HPGe detectors, as compared to LaBr$_3$:Ce either with a PMT or a SiPM array readout, is clear. However, the much longer signal fall time of HpGe as compared to LaBr$_3$: Ce has prompted the choice of these last detectors in the FAMU experiment, to have a better separation between signal (delayed X-rays) and background (prompt X-rays).

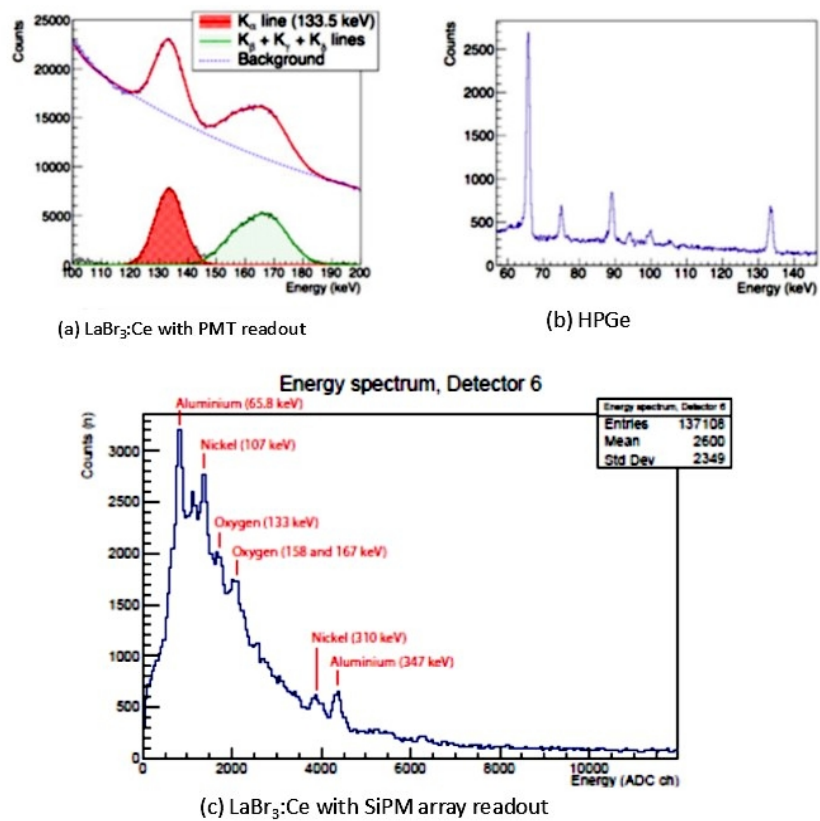

**Figure 5.** Muonic X-ray spectra recorded in the FAMU experiment at RAL using (a) 1" LaBr3:Ce counters with PMT readout, $K_\beta$ and $K_\gamma$ lines are not resolved; (b) the HPGe detector and (c) 1/2" LaBr3:Ce counters with SiPM array readout, from reference (Bonesini et al., 2019)

For nuclear spectroscopy at higher energy, CdTe and CdZnTe (CZT) have found increasing applications (Squillante and Entine, 1992), (McConnel et al., 2000), (Zambelli et al., 2020). Their main advantage is to operate at room temperature, with no need for cooling and to have a high count-rate capability (up to $10^8$ photons per second per mm$^2$). The FWHM energy resolution of CZT is better than the one of any scintillating crystal on the market. Values around 1.7 % (5 %) at 662 (122) keV are within reach. In addition these detectors are not hygroscopic. CZT detectors may be built either as a single crystal up to a size 20 mm square or as an array made of smaller elements. The top and bottom sides are metalized with the cathode and anode terminals plus any guard rings to control electrical parameters and sensitivity. As CZT crystals are extremely brittle, low stress designs to produce reliable assemblies are required. Adding Selenium to the CZT matrix, an outstanding resolution: up to 0.87 % at 662 keV and 4.6 % at 81 keV is reached for detectors of a size of $4.5 \times 4.5 \times 10.8$ mm$^3$ (Roy et al., 2019).

## 3.2 Future improvements

Future improvements are mainly connected to the development of non-hygroscopic crystals, to the improvement of LaBr$_3$: Ce characteristics with Li$^+$, Na$^+$,Mg$^{2+}$, Ca$^{2+}$, Sr$^{2+}$ or Ba$^{2+}$ co-doping (Yang et al., 2012), (Alekhin et al., 2013) to increase energy resolution and the development of new crystals responsive to both gamma rays and neutrons, such as CLBBC.

As shown in reference (Yang et al., 2012), with a Sr or Ba co-doping there is a light output improvement of $\sim$ 25 %, reducing the energy FWHM resolution of about 14 % at 120 keV and 662 keV in a 60 mm round, 80 mm thick crystal. The decay time increases only slightly: from 17.2 ns to 18.2 ns ( with Sr co-doping) and to 19.1 ns (with Ba co-doping). At the present moment, the best compromise is still LaBr$_3$: Ce crystals that have an excellent energy resolution (up to 2.5 % at 662 keV), good timing properties: a 100 ps coincidence resolving time was obtained, at the price of a moderate intrinsic activity ($\sim$ 0.2 Bq/g) due to the presence of $^{138}$La. Detectors up to 3.5" have been developed and 1" round detectors are in common use. They have extensive applications from PET or TOF PET to ones as a satellite payload in a harsh environment. For their readout, the new wave is the SiPM based one, which is insensitive to magnetic fields and requires limited power to work at the cost of a temperature dependence of gain.

## 4 EXAMPLES OF CRYSTAL-BASED GAMMA RAY DETECTION SYSTEMS

In the following some gamma-ray detection systems, based on scintillating crystals, are discussed. They cover the field from fundamental physics: the FAMU experiment at RAL for the study of hyperfine spectroscopy of muonic hydrogen and the NUMEN experiment at INFN LNS, to study $0\nu\beta\beta$ decay, to astrophysics: the Academy of China GECAM observatory for X-ray bursts. All detectors involve LaBr$_3$: Ce detectors in a form or the other and suffer from different experimental problems and challenges.

For the FAMU experiment at RAL, a fast detectors response (fall time $\leq$ 200 ns) is required to separate the prompt background from the delayed X-ray signal. Instead in G-NUMEN, a high-rate capability (up to 300 kHz) is needed, while in the GECAM observatory the radiation damage to SiPM arrays, due to cosmic high-energy protons, is an issue. Problems in all the examples are enhanced by the common requirement to have large area detectors: at least 1" in size.

## 4.1 The FAMU apparatus

The FAMU (Fisica degli Atomi Muonici) experiment at RIKEN-RAL (Pizzolotto et al., 2020), (Adamzack et al., 2018) aims at high-precision spectroscopic studies of muonic hydrogen. In particular to measure the hyperfine splitting $\Delta E^{hfs}$ in the 1S state of muonic hydrogen (Bakalov et al., 1993), (Adamczak et al., 2012), (Vacchi et al., 2012). It makes use of a high-intensity pulsed low-energy muon beam (Matsuzaki et al., 2001), stopping in a hydrogen target, to produce muonic hydrogen (in a mixture of singlet F=0 and triplet F=1 states) and a tunable mid-IR (MIR) pulsed high-power laser (Baruzzo et al., 2024) to excite the hyperfine splitting (HFS) transition of the 1S muonic hydrogen (from F=0 to F=1 states). Exploiting the muon transfer from muonic hydrogen to another higher-Z gas in the target (such as $O_2$ or Ar), the $(\mu^- p)_{1S}$ HFS transition will be recognized by an increase in the number of X-rays from the $(\mu Z^*)$ cascade, while tuning the laser frequency $\nu_0$ ($\Delta E_{HFS} = h\nu_0$). From the measure of $\Delta E^{hfs}(\mu^- p)_{1S}$ the Zemach radius $r_Z$ of the proton (Zemach, 1956) may be deduced with a precision better than $10^{-2}$, thus shedding new light on the problem of the proton radius puzzle (Antognini et al., 2013), (Pohl et al., 2010).

The signal X-ray detection (around 130 keV) is based on LaBr$_3$: Ce crystals read either by photomultipliers (Baldazzi et al., 2017) or SiPM arrays (Bonesini et al., 2020) [7]. One HPGe detector is used for inter-calibration. In this experiment, a fast detector response (fall times $\leq$ 200 ns) is needed to separate signal X-rays from the background.

The FAMU setup for the 2023-2024 data taking is based on one ORTEC GEM-S5020P4 HpGe for inter-calibrations, and 34 LaBr$_3$: Ce detectors:

- six 1" round, 1" thick detectors are read by conventional PMTs (Baldazzi et al., 2017)
- sixteen 1" round, 0.5" thick detectors are read by SiPM arrays (Bonesini et al., 2023a)
- twelve 1/2" cubic detectors are read by SiPM arrays (Bonesini et al., 2016)

In the 2024 data taking the twelve 1/2" detectors were replaced by 1" round, 0.5" thick detectors. An enlargement of layout of the FAMU experimental setup in the region where X-ray detectors are placed is shown in Figure 6, where the 34 LaBr$_3$: Ce detectors are arranged in three crowns.

The detectors with a PMT readout have a fully active divider and a custom Digital Pulse Processor (DPP) based on 12-bit 500 Ms/s Analog Devices ADC, as explained in reference (Baldazzi et al., 2017). Timing and FWHM energy resolution of the 3 types of detectors are shown in table 8. FWHM energy resolutions at $^{137}$Cs and $^{57}$Co peaks are from laboratory measurements, while at 142 keV (Ag peak) they are from beam data at RAL with 55 Mev/c impinging muons. For comparison, the FWHM energy resolution at the 142 keV muonic silver peak is $1.26 \pm 0.17\%$ from the HPGe detector, at the cost of a much longer fall time. Rise time and fall time are 10-90 % and are measured in laboratory at the $^{137}$Cs peak. While 1/2" detectors with a SiPM readout use a conventional parallel ganging, the 1" detectors use the 4-1 NI layout, which reduce fall time by a factor 2 at least.

**Table 8.** Average performances of FAMU LaBr$_3$: Ce detectors. Rise time and fall time (10-90 %) refer to detectors' analog outputs.

| det. type | rise time (ns) | fall time (ns) | R(%) @ $^{137}$Cs | R(%) @ $^{57}$Co | R(%) @ 142 keV |
|---|---|---|---|---|---|
| 1" - PMT | 14 $\pm$ 1 | $\sim$ 60 | 3.5 - 4.6 | 7.2-8.1 | 11.5 $\pm$ 0.2 |
| 1" - SiPM | 29.3 $\pm$ 1.5 | 147.1 $\pm$ 12.8 | 2.94 $\pm$ 0.14 | 8.03 $\pm$ 0.39 | 8.2 $\pm$ 0.7 |
| 1/2" - SiPM | 42.8 $\pm$ 1.5 | 372.4 $\pm$ 17.4 | 3.27 $\pm$ 0.11 | 8.44 $\pm$ 0.63 | 7.5 $\pm$ 0.3 |

---

[7] A preliminary study to asses if non-hygroscopic crystals, such as PrLuAg and Ce: GAAG, may be suitable was done, with a negative response, and is reported in reference (Bonesini et al., 2017)

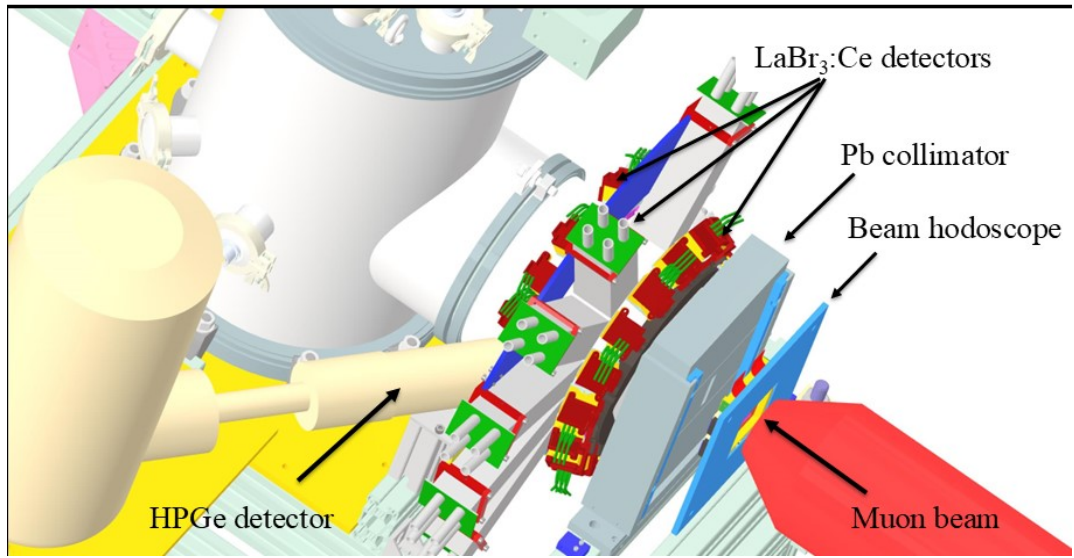

**Figure 6.** Enlargement of the FAMU experimental setup in the region of X-rays detectors. The three crowns of LaBr3:Ce detectors are shown, together with the HPGe detector.

### 4.2 The G-NUMEN apparatus

The G-NUMEN LaBr$_3$:Ce array is the future gamma spectrometer for the NUMEN experiment at INFN-LNS (Cappuzzello et al., 2023), that aims to study neutrinoless double beta decay ($0\nu\beta\beta$). Informations on nuclear matrix elements of the $0\nu\beta\beta$ decay will be obtained through double-exchange (DCE) reactions generated by heavy ions. As cross sections of the order of a few nb are expected for DCE reactions, an apparatus with high sensitivity and resolution is needed. The experimental apparatus is made of the high-acceptance spectrometer (MAGNEX), a focal plane detector (FPD) and the G-NUMEN array. The G-NUMEN array include 110 LaBr3:Ce detectors placed around the scattering chamber, as shown in figure 7, and will be used to detect the characteristic gamma-ray transitions in DCE events. The used detectors will have to sustain a rate up to 300 kHz per crystal. Having a conventional PMT readout with Hamamatsu R6231 PMTs, phototube stability under high rates is a relevant issue, together with linearity. Used LaBr3:Ce crystals are 1.5" round, 2" thick. At the $^{137}$Cs peak, a FWHM energy resolution $\sim 3\ \%$ has been obtained.

### 4.3 The GECAM experiment

The Gravitational Wave High-Energy Electromagnetic Counterpart All-Sky Monitor (GECAM) is a Chinese Academy of Sciences project aiming at the detection of the high-energy counterparts of gravitational waves (Feng and Su, 2024). Scientific goals of GECAM include the detection of gamma-ray bursts (GRBs), solar flares (SGLs), and fast radio bursts (FRBs). About 300 GRBs have been detected by GECAM, including the brightest GDR of all time: the GRB221009A. This GDR was detected also by many other instruments worldwide, but GECAM was the only one not suffering from signal saturation and pulse pile-up.

GeCAM includes four instruments: GECAM A/B, GECAM C and GECAM D, launched between December 2020 and March 2024, equipped with Gamma-ray detectors (GDRs) mainly based on 3" round,

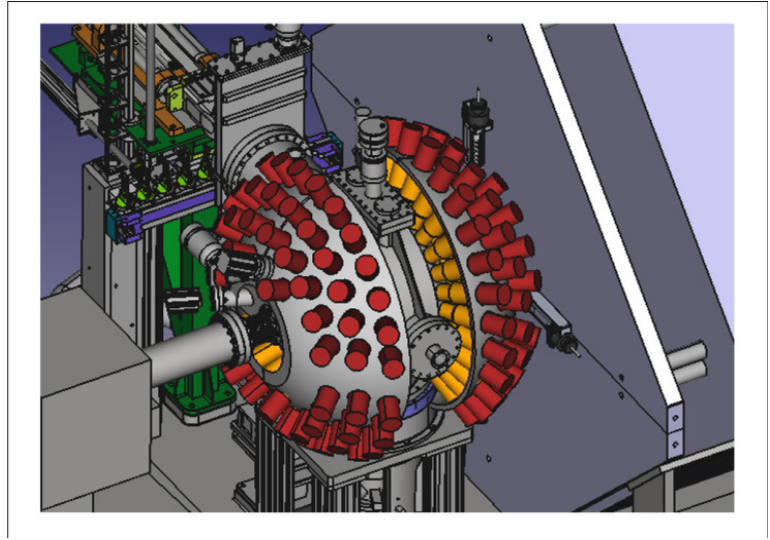

**Figure 7.** Image of the NUMEN experimental setup in the region of X-rays detectors, from reference (Cappuzzello et al., 2023).

15 mm thick Labr$_3$:Ce detectors. The detectors have readout based on SiPM: 64 $6 \times 6$ mm$^2$ SiPM arranged in a circular shape (He et al., 2023). A FWHM energy resolution $\sim 5.3$ % was reached at the $^{137}$Cs peak in laboratory tests, matching the target requirement of 8 %. Energy resolution and linearity were evaluated using radioactive sources and a high-energy X-ray calibration facility (HCXF).

In-orbit gain variation of SiPMs, due to temperature excursions, was corrected reducing non-uniformity from 17 % to 0.6 %. Due to the presence of background cosmic high-energy protons an increase in the dark currents of SiPM $\sim 0.43 \mu$A/day were observed, connected to the displacement damage in SiPM under irradiation. To circumvent this phenomenon, an in-situ current annealing was studied (Gu et al., 2023). The generated local heat, during current flow, may repair damage and defects.

## 5 CONCLUSIONS

The field of high energy X-ray and gamma-ray detectors based on inorganic scintillating crystals is in continuous evolution. The last years have seen a relevant progress in the development of new scintillating crystals. Energy resolutions up to 2 % at 662 keV are within reach and heavy and fast scintillators have been produced. The present development effort is mainly concentrated on the improvement of LaBr$_3$: Ce crystals with co-doping, the introduction of new non-hygroscopic crystals and the development of either new PMTs with higher photocathode QE or novel SiPM arrays with reduced dark noise, lower bias voltages and larger dimensions. In addition efforts for the development of new ASICs for the readout of segmented scintillating crystal are under way. The readout of scintillating inorganic crystals with SiPM is promising and FWHM energy resolutions and timing are reaching what is obtained with conventional PMTs.

However a full understanding of the finite measured energy resolution of scintillating crystals (below the intrinsic one) has not yet been fully reached, even if there are hints that non-linearity in the crystal's response or non-uniformity may be the reason.

# CONFLICT OF INTEREST STATEMENT

The authors declare that the research was conducted in the absence of any commercial or financial relationships that could be construed as a potential conflict of interest.

# AUTHOR CONTRIBUTIONS

The author confirms sole responsibility for the following: study conception and design, data collection, analysis and interpretation of results, and manuscript preparation.

# FUNDING

This research was funded by INFN Commissione 3, inside the 2024 funding for the FAMU experiment at RAL.

# ACKNOWLEDGMENTS

This review is the outcome of a ten years old research activity in the FAMU experiment at RAL, where a lot of the shown problems were tackled. I would like to thank all my colleagues of the FAMU collaboration, in particular Andrea Vacchi, Roberto Bertoni,Ludovico Tortora and Emiliano Mocchiutti for many interesting discussions and suggestions. In addition I would like to thank Francesco Caponio and Andrea Abba from Nuclear Instruments srl for many interesting discussions on SiPM readout and M. Saviozzi from CAEN srl for a lot of help on electronics issues.

# SUPPLEMENTAL DATA

No supplemental data are available

# DATA AVAILABILITY STATEMENT

No specific datasets for this study are available.

*M. Bonesini*  **Gamma-ray and high-energy X-ray detection with scintillating crystals**Kuhn et al. (2006). Performance assessment of pixelated LaBr$_3$ detector modules for TOF PET. *IEEE Trans. Nucl. Sci.* 53, 1090–1095. doi:10.1109/tns.2006.873708

Matsuzaki, T. et al. (2001). The RIKEN RAL pulsed muon facility. *Nucl. Instr. Meth.* A465, 365. doi:10.1016/s0168-9002(01)00694-5

McConnel, M. et al. (2000). Three-dimensional imaging and detection efficiency performance of orthogonal CZT strip detectors. *Proceedings of SPIE* 4141, 157–167. doi:10.1117/12.407576

Moon, R. (1948). Inorganic crystals for the detection of high energy particles and quanta. *Phys. ReV.* 73, 1210

Moses, W. (2007). Recent advances and future advances in time-of-flight PET. *Nucl. Instr. Meth.* A580, 919–924. doi:10.1016/j.nima.2007.06.038

Moses, W. and Shah, K. (2005). Potential of RbGd$_2$Br$_7$:Ce, LaBr$_3$:Ce and LuI$_3$:Ce in nuclear medical imaging. *Nucl. Instr. Meth.* 537, 317. doi:10.1016/j.nima.2004.08.034

Moszynski, M. et al. (2002). Intrinsic energy resolution of NaI (Tl). *Nucl. Instr. Meth.* A484, 259–269. doi:10.1016/S0168-9002(01)01964-7

Moszynski, M. et al. (2008). A comparative study with PMTs, Avalanche Photodiodes, Photodiodes and PIN photodiodes in Gamma Spectroscopy with LaBr$_3$ Crystals . *IEEE Trans. Nucl. Sci.* Berkeley. doi:10.1109/tns.2008.2005110

Nikl, M. (2006). Scintillator detectors for X-rays. *Meas. Science and Technol.* 17, R37–R54. doi:doi: 10.1088/0957-0233/17/4/r01

Nikl, M. et al. (1996). Slow components in the photoluminescence and scintillation decay of PbWO$_4$ single crystals. *Phys. Status Solifi* b195, 311–323

Ogawa, M. (2016). MEG II Collaboration, Master Thesis, University of Tokio

Omer, M. et al. (2013). Performance of LaBr3(Ce) array detector system for non-destructive inspection of special nuclear material by using nuclear resonance fluorescence. *IEEE International Conference on Technologies for Homeland Security (HST), Waltham, MA, USA* , 671–676doi:10.1109/THS.2013.6699084

Otte, A. et al. (2017). Characterization of three high efficiency and blue sensitive silicon photomultipliers. *Nucl. Instr. Meth.* A846, 106

Pani, R. et al. (2008). Gamma-ray spectroscopy with LaBr3:Ce scintillation crystal coupled to an ultra high quantum efficiency pmt. *Nuclear Science Symposium* , 2462 – 2466doi:10.1109/nssmic.2008.4774853

Pascu, S. et al. (2024). New readout system of the FATIMA detectors based on silicon photomotipliers arrays. *Nucl. Instr. Meth.* A17001. doi:10.1016/j.nima.2024.170001

Pizzolotto, C. et al. (2020). The FAMU experiment: Muonic hydrogen high precision spectroscopy studies. *Eur. Phys. J. A* 7, 185. doi:10.1140/epja/s10050-020-00195-9

Pohl, R. (1938). Zusammen fassender bericht uber elektronenleitung und photochemische vorgang in alkalihalogeneid kristallen. *Zeit. Physik* 39, 36

Pohl, R. et al. (2010). The size of the proton. *Nature* 466, 213–216. doi:10.1038/nature09250

Poleshchuck, O. et al. (2021). Performances tests of a LaBr3:Ce detector coupled to a SiPM array and the GET electronics for $\gamma$-ray spectroscopy in a strong magnetic field. *Nucl. Istr. Meth.* A987, 1648663. doi:10.1016/j.nima.2020.164863

Quarati, F. et al. (2007). X-ray and gamma-ray response of a 2" x 2" LaBr$_3$: Ce scintillation detector . *Nucl. Instr. Meth.* A574, 115–120. doi:10.1016/j.nima.2007.01.161

Quarati, F. et al. (2012). Scintillation and detection characteristics of high-sensitivity CeBr$_3$ gamma-ray spectrometers. *Nucl. Instr. Meth.* A729, 596. doi:10.1016/j.nima.2013.08.005

Roentgen, W. (1896). On a new kind of rays. *Science* 3 (59), 227–230
**Frontiers** 24